\documentclass[sigconf]{acmart}

\usepackage{amsmath}
\usepackage{subcaption}
\usepackage{bm}

\usepackage{algorithm}
\usepackage{algorithmic}
\usepackage{graphicx} % Required for inserting images

\AtBeginDocument{%
  \providecommand\BibTeX{{%
    \normalfont B\kern-0.5em{\scshape i\kern-0.25em b}\kern-0.8em\TeX}}}

\setcopyright{acmlicensed}
\copyrightyear{2024}
\acmYear{2024}
\acmDOI{XXXXXXX.XXXXXXX}

\acmConference[KDD'24]{30th ACM SIGKDD Conference on Knowledge Discovery and Data Mining}{August 25--29,
  2024}{Barcelona, Spain}
\acmISBN{978-1-4503-XXXX-X/18/06}

\begin{document}

%%
%% The "title" command has an optional parameter,
%% allowing the author to define a "short title" to be used in page headers.
\title{Adaptive Reinforcement Learning for Dynamic Configuration Allocation in Pre-Production Testing
}

%%
%% The "author" command and its associated commands are used to define
%% the authors and their affiliations.
%% Of note is the shared affiliation of the first two authors, and the
%% "authornote" and "authornotemark" commands
%% used to denote shared contribution to the research.
\author{Yu Zhu}
\affiliation{%
  \institution{University of California, Santa Cruz}
  \city{Santa Cruz}
  \state{CA}
  \country{USA}
}
\email{yzhu201@ucsc.edu}

%\author{Rohit Pandey}
%\affiliation{%
%  \institution{Microsoft}
%\city{Redmond}
%  \state{WA}
%  \country{USA}
%}
%\email{ropandey@microsoft.com}

%\author{Akshay Sathiya}
%\affiliation{%
%  \institution{Microsoft}
% \city{Redmond}
%  \state{WA}
%  \country{USA}
%}
%\email{asathiya@microsoft.com}

% \author{Huifen Chan}
% \affiliation{%
%  \institution{Tsinghua University}
%  \streetaddress{30 Shuangqing Rd}
%  \city{Haidian Qu}
%  \state{Beijing Shi}
%  \country{China}}

%%
%% By default, the full list of authors will be used in the page
%% headers. Often, this list is too long, and will overlap
%% other information printed in the page headers. This command allows
%% the author to define a more concise list
%% of authors' names for this purpose.
\renewcommand{\shortauthors}{Zhu, et al.}

%%
%% The abstract is a short summary of the work to be presented in the
%% article.
\begin{abstract}
Ensuring reliability in modern software systems requires rigorous pre-production testing across highly heterogeneous and evolving environments. Because exhaustive evaluation is infeasible, practitioners must decide how to allocate limited testing resources across configurations where failure probabilities may drift over time. Existing combinatorial optimization approaches are static, ad hoc, and poorly suited to such non-stationary settings. We introduce a novel reinforcement learning (RL) framework that recasts configuration allocation as a sequential decision-making problem. Our method is the first to integrate Q-learning with a hybrid reward design that fuses simulated outcomes and real-time feedback, enabling both sample efficiency and robustness. In addition, we develop an adaptive online–offline training scheme that allows the agent to quickly track abrupt probability shifts while maintaining long-run stability. Extensive simulation studies demonstrate that our approach consistently outperforms static and optimization-based baselines, approaching oracle performance. This work establishes RL as a powerful new paradigm for adaptive configuration allocation, advancing beyond traditional methods and offering broad applicability to dynamic testing and resource scheduling domains.
\end{abstract}

%%
%% The code below is generated by the tool at http://dl.acm.org/ccs.cfm.
%% Please copy and paste the code instead of the example below.
%%
\begin{CCSXML}
<ccs2012>
<concept>
<concept_id>10002950.10003648.10003649.10003655</concept_id>
<concept_desc>Mathematics of computing~Causal networks</concept_desc>
<concept_significance>500</concept_significance>
</concept>
</ccs2012>
\end{CCSXML}

\ccsdesc[500]{Mathematics of computing~Reinforcement Learning}

%%
%% Keywords. The author(s) should pick words that accurately describe
%% the work being presented. Separate the keywords with commas.
\keywords{A/B testing, Reinforcement Learning}

\received{20 February 2007}
\received[revised]{12 March 2009}
\received[accepted]{5 June 2009}

%%
%% This command processes the author and affiliation and title
%% information and builds the first part of the formatted document.
\maketitle

\section{Introduction}
Modern software systems are deployed across increasingly diverse and heterogeneous environments. Variability arises from hardware platforms, operating system versions, virtualization technologies, and execution contexts, each of which can introduce distinct performance characteristics and potential sources of instability. This heterogeneity poses a major challenge for pre-production testing pipelines, which aim to identify reliability and performance issues before new software versions are deployed at scale. If underrepresented configurations fail in production, the consequences may include service interruptions, degraded user experience, or costly rollbacks. Thus, systematic and adaptive methods for test allocation are critical to ensuring robust system reliability.

A central difficulty stems from the infeasibility of exhaustively testing all possible environment configurations. Even modest settings with tens of hardware types, multiple operating systems, and virtualization options can yield thousands of possible combinations. Given resource constraints, practitioners typically evaluate only a subset of configurations during A/B testing or canary deployments. The challenge, therefore, is to allocate a limited testing budget across configurations so as to maximize the likelihood of detecting potential failures, thereby improving the statistical power and reliability of pre-production evaluations \cite{nie2011,kuhn2013}.

Traditionally, this allocation problem has been addressed using combinatorial optimization (CO) techniques, such as simulated annealing, greedy heuristics, or integer programming \cite{schrijver2003}. These methods attempt to balance coverage across dimensions by constructing representative subsets of configurations. However, they suffer from three key limitations:

\begin{enumerate}
    \item \textbf{Static assumptions.} Many optimization methods assume that configuration failure probabilities are fixed. In practice, environments are dynamic: probabilities of encountering errors shift due to hardware degradation, software patches, or workload changes. Static allocation strategies quickly become obsolete in such settings.
    \item \textbf{Ad-hoc parameterization.} Cost functions and hyperparameters are often tuned manually. For instance, simulated annealing relies on temperature schedules and acceptance rates, which may not generalize well across environments. This limits reproducibility and robustness \cite{bergstra2012}.
    \item \textbf{Lack of feedback integration.} Traditional CO strategies typically ignore real-time testing feedback. Once a schedule is constructed, allocations remain fixed even as outcomes accumulate. This leads to inefficiencies, especially in scenarios where early signals indicate misallocation of testing resources.
\end{enumerate}

Reinforcement Learning (RL) provides a natural framework for sequential decision-making under uncertainty \cite{sutton2018}. By treating configuration allocation as an RL problem, an agent can iteratively adjust allocations based on observed signals, thereby improving coverage over time. Unlike static optimization, RL explicitly models the exploration--exploitation trade-off: the agent explores new configurations to gather information, while exploiting known high-risk configurations to ensure adequate testing.

Recent research has demonstrated RL’s promise in related domains. In \textit{combinatorial optimization}, RL has been applied to routing, resource scheduling, and job-shop problems \cite{bello2016,mao2016}, often outperforming handcrafted heuristics. In \textit{adaptive experiment design}, RL has been shown to improve statistical efficiency by sequentially refining treatment allocations based on observed outcomes \cite{Shen_2025}. In \textit{systems optimization}, deep RL has enabled dynamic scaling and load balancing in cloud platforms \cite{mao2016}. Collectively, these results suggest that RL is well-suited for adaptive allocation in heterogeneous testing environments.

Despite its promise, applying RL in this setting raises several challenges:

\paragraph{1. Limited and costly feedback.} In real-world testing pipelines, collecting outcome data (e.g., failure signals) can be slow, expensive, or risky. Direct training of RL agents on live systems is impractical. To address this, hybrid approaches have been proposed that combine simulated and real-world experiences. Reward shaping \cite{ng1999,grzes2017}, transfer learning \cite{pan2010}, and offline reinforcement learning with experience replay \cite{mnih2015} are examples of techniques that reduce dependence on costly online feedback.

\paragraph{2. Non-stationary environments.} Failure probabilities are not fixed. Configuration risks may shift abruptly due to software updates, workload spikes, or latent factors. RL agents trained under stationary assumptions may underperform in practice. Addressing non-stationarity requires adaptive strategies such as meta-learning \cite{finn2017}, adaptive exploration \cite{zhu2022adaptive}, and policy adaptation to concept drift \cite{yu2020}.

\paragraph{3. High-dimensional allocation space.} Even when focusing on a single dimension (e.g., hardware type), the number of possible allocations grows combinatorially with the number of configurations and test units. Direct enumeration is infeasible. Scalable RL methods, such as function approximation and state aggregation, are therefore necessary.

In this paper, we introduce a reinforcement learning framework for adaptive configuration allocation in pre-production testing. Our work makes the following contributions:

\begin{enumerate}
    \item \textbf{Problem formalization.} We formalize the allocation of testing units across heterogeneous configurations as a sequential decision-making problem. The agent seeks to maximize the number of configurations achieving a signal detection threshold, reflecting coverage objectives critical for robust testing.
    \item \textbf{RL-based allocation strategy.} We propose a Q-learning framework that integrates simulated outcomes (based on historical estimates) with observed outcomes, enabling robust updates under limited feedback. Our reward function is carefully shaped to balance statistical coverage with real-world efficiency.
    \item \textbf{Adaptation to non-stationarity.} We develop mechanisms to handle abrupt changes in configuration probabilities. By combining online updates with offline simulations, our approach adapts quickly to drift and maintains stable coverage.
    \item \textbf{Empirical evaluation.} Through simulation studies, we compare static optimization, rolling Lagrangian methods, and our RL approach under dynamic environments. Results show that RL achieves superior adaptability and performance, approaching the oracle benchmark.
\end{enumerate}

The remainder of the paper is organized as follows. Section~\ref{sec:problem} presents the problem formulation and constraints. Section~\ref{sec:methodology} details the reinforcement learning methodology, including the state, action, reward, and update mechanisms. Section~\ref{sec:simulation} describes the simulation setup, allocation strategies, and evaluation metrics. Section~\ref{sec:simulation} reports experimental results, including statistical comparisons. Section~\ref{sec:conclusion} concludes with directions for future research.

\section{Problem Formulation}
\label{sec:problem}

\subsection{Pre-Production Configuration Allocation}

Consider a pre-production testing environment where a limited budget of $N$ testing units (e.g., virtual instances, devices, or simulated runs) must be allocated across $C$ configuration types. Each configuration type, denoted $c_i$ for $i=1,\ldots,C$, may represent a specific combination of system attributes such as hardware model, virtualization setting, and operating system version. The objective is to allocate units across these types in order to maximize the likelihood of detecting potential failures prior to deployment.

\begin{figure*}
    \centering
    \includegraphics[width=0.7\textwidth]
{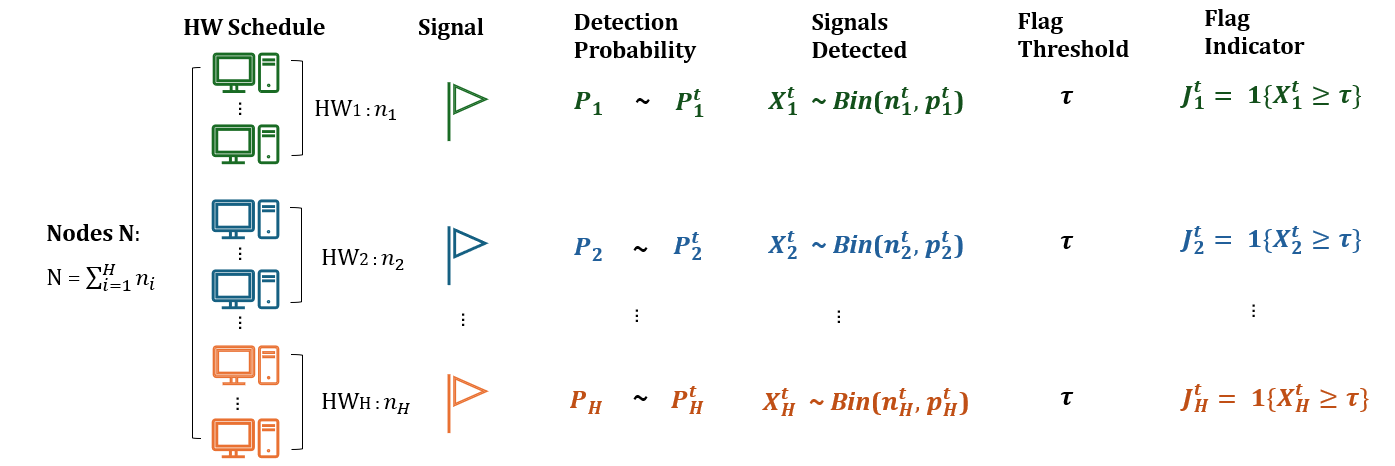}
    \caption{Pre-production configuration allocation problem description.}
    \label{fig:problem_des}
\end{figure*}

This can be represented as follows (as in Figure \ref{fig:problem_des}). Formally, an allocation at time $t$ is defined by the vector
\[
S_t = \big[n_1(t),\,n_2(t),\,\ldots,\,n_C(t)\big],
\]
where $n_i(t)$ is the number of units assigned to configuration type $c_i$ at time $t$, subject to the budget constraint
\[
\sum_{i=1}^C n_i(t) = N, \quad n_i(t)\in\mathbb{Z}^{+}.
\]
The allocation space grows combinatorially with $C$ and $N$, making exhaustive search infeasible for realistic values.

For each configuration type $c_i$, let $p_i(t)$ denote the probability of detecting a failure (or ``signal'') when a single unit is allocated at time $t$. If $n_i(t)$ units are assigned to type $c_i$, the number of detected signals follows a binomial distribution:
\[
X_i(t) \sim \text{Binomial}\big(n_i(t), p_i(t)\big).
\]
The testing process is said to \emph{cover} configuration $c_i$ if at least one signal is observed:
\[
J_i(t) = \mathbf{1}\{X_i(t) \geq 1\}.
\]
The overall coverage at time $t$ is then
\[
D_t = \sum_{i=1}^{C} J_i(t).
\]
Maximizing $D_t$ ensures that as many configuration types as possible are represented by at least one detected signal, aligning with the goal of broad reliability assurance. This formulation is closely related to objectives studied in group testing and adaptive experiment design \cite{aldous1989, nie2011}.

Several constraints complicate the allocation problem:

\begin{enumerate}
    \item \textbf{Finite budget.} Only $N$ units can be deployed in each testing cycle, imposing a strict resource constraint.
    \item \textbf{Unobserved probabilities.} The true $p_i(t)$ are unknown and must be estimated from historical or recent data, leading to statistical uncertainty.
    \item \textbf{Non-stationarity.} Probabilities $p_i(t)$ may vary with $t$, reflecting evolving environments. Static allocation policies become suboptimal under such dynamics \cite{garivier2011, besbes2014}.
    \item \textbf{Coverage trade-offs.} Allocating more units to high-risk configurations increases detection probability but reduces diversity, while allocating thinly across many configurations risks missing critical failures. The allocation must strike a balance.
\end{enumerate}

These challenges distinguish the problem from classical combinatorial testing \cite{kuhn2013} and motivate adaptive strategies.

\subsection{Sequential Decision-Making}

The allocation can be posed as an optimization problem:
\[
\max_{n_1(t), \ldots, n_C(t)} \; \mathbb{E}[D_t] \quad 
\text{s.t.} \quad \sum_{i=1}^C n_i(t) = N, \; n_i(t)\in\mathbb{Z}^{+}.
\]
Since
\[
\mathbb{E}[J_i(t)] = 1-(1-p_i(t))^{n_i(t)},
\]
the expected coverage is
\[
\mathbb{E}[D_t] = \sum_{i=1}^{C} \left[ 1 - (1-p_i(t))^{n_i(t)} \right].
\]
This nonlinear objective resembles problems studied in stochastic optimization, subset selection, and multi-armed bandits \cite{audibert2009,bubeck2012}. However, unlike standard bandit settings, the allocation is high-dimensional (simultaneous allocation across $C$ types) and must adapt to changing $p_i(t)$ values over time.

Given the temporal evolution of probabilities and feedback accumulation, the allocation problem is better viewed as a sequential decision-making process. At each step $t$:
\begin{enumerate}
    \item The agent observes historical signals and forms estimates $\hat{p}_i(t)$ for each type $i$.
    \item An allocation $S_t$ is chosen subject to the budget constraint.
    \item Signals $X_i(t)$ are observed, providing feedback for updating estimates.
    \item The process repeats for $t+1$.
\end{enumerate}

This sequential structure aligns naturally with a reinforcement learning framework. The state corresponds to the current allocation and estimated probabilities; the action corresponds to a reallocation decision; and the reward is the observed coverage $D_t$. The challenge is to design an adaptive policy $\pi$ that maximizes long-run coverage:
\[
\pi^* = \arg\max_\pi \; \mathbb{E}\left[\sum_{t=1}^T D_t \,\big|\, \pi \right].
\]

\section{Methodology}
\label{sec:methodology}

The configuration allocation problem described in Section~\ref{sec:problem} can be framed as a sequential decision-making task under uncertainty. At each time step, an agent must decide how to allocate limited testing units across heterogeneous configuration types, observe outcomes in the form of detected signals, and update its strategy accordingly. The environment is dynamic, with probabilities of signal detection varying over time, making static optimization insufficient. Reinforcement Learning (RL) provides a principled approach for adaptively improving allocation policies through interaction with such environments \cite{sutton2018}.

Among RL algorithms, Q-learning is particularly well-suited to this problem. It is a model-free method that does not require explicit knowledge of transition dynamics, which are unknown in practice. Moreover, Q-learning is flexible in handling large and combinatorial state-action spaces when combined with function approximation or structured exploration. Importantly, Q-learning can incorporate both simulated and real feedback, making it an appropriate choice when observations are sparse or costly.

\subsection{Q-learning Framework for Configuration Allocation}

We formulate the allocation task as a Markov decision process (MDP) with state, action, and reward components defined as follows.

\subsubsection{State Space}
The state at time $t$, denoted $S_t$, captures both the allocation and the estimated signal detection probabilities:
\[
S_t = \big[n_1(t), \ldots, n_C(t), \hat{p}_1(t), \ldots, \hat{p}_C(t)\big].
\]
Here $n_i(t)$ is the number of units assigned to configuration $c_i$, and $\hat{p}_i(t)$ is the estimated probability of detection, computed from historical data as described in Section~\ref{sec:simulation}. This representation allows the agent to reason jointly about allocation and estimated risk.

\subsubsection{Action Space}
An action $A_t$ is defined as a reallocation of testing units between configuration types:
\[
A_t = (i,j,\Delta),
\]
where $\Delta$ units are reallocated from configuration $c_i$ to $c_j$, subject to feasibility constraints ($n_i(t)\geq \Delta$, $n_j(t)+\Delta \leq N$). This flexible structure captures the full range of incremental reallocations, from small adjustments to larger redistributions.

\subsubsection{Reward Function}
The reward at time $t$ is designed to reflect the coverage objective: maximizing the number of configuration types that achieve at least one detected signal. Directly using $D_t$ as the reward is possible, but it introduces high variance due to stochastic binomial outcomes. To stabilize learning, we adopt a hybrid reward-shaping approach \cite{ng1999, grzes2017}:
\[
R_t = \sum_{i=1}^{C} \Big[ \omega_1 \cdot \mathbf{1}\{x_{i,t}\geq \tau\} + \omega_2 \cdot \mathbf{1}\{X_i(t)\geq \tau\} \Big],
\]
where $x_{i,t}$ are simulated outcomes based on estimated probabilities $\hat{p}_i(t)$, $X_i(t)$ are observed signals, $\tau$ is the detection threshold (typically $\tau=1$), and $\omega_1,\omega_2$ are weights balancing simulated versus observed contributions. This structure enables the agent to pre-train on simulated signals while still grounding learning in real feedback, reducing sample complexity.

\subsubsection{Q-value Update}
Q-learning updates state-action values using the Bellman equation:
\[
Q(S_t,A_t) \leftarrow Q(S_t,A_t) + \alpha \big[R_t + \gamma \max_{A'} Q(S_{t+1},A') - Q(S_t,A_t)\big],
\]
where $\alpha$ is the learning rate and $\gamma$ is the discount factor. Over time, $Q$ converges to the expected cumulative reward of taking action $A_t$ in state $S_t$ under the optimal policy.

\subsubsection{Exploration vs. Exploitation}
To balance exploration of new allocations with exploitation of high-value actions, we employ an $\epsilon$-greedy strategy. With probability $\epsilon$, a random feasible reallocation is chosen; otherwise, the agent selects the action maximizing $Q(S_t,A_t)$. To adapt over time, $\epsilon$ decays gradually as the agent gains confidence. Under abrupt shifts in the environment, $\epsilon$ may be temporarily increased to encourage renewed exploration.

\subsection{Adapting to Dynamic Environments}

A key challenge is that detection probabilities $p_i(t)$ evolve over time, reflecting non-stationary environments \cite{garivier2011, besbes2014}. Standard Q-learning, which assumes stationary dynamics, may converge to outdated policies. We introduce two mechanisms to improve adaptability:

\begin{enumerate}
    \item \textbf{Hybrid online/offline learning.} The agent updates policies daily with observed outcomes (online learning), while simultaneously generating offline simulations using the latest estimates $\hat{p}_i(t)$ (offline pre-training). This mirrors experience replay \cite{mnih2015} and accelerates adaptation by exposing the agent to a richer set of trajectories.
    \item \textbf{Adaptive exploration.} When sudden probability shifts are detected (e.g., significant deviation between observed $X_i(t)$ and expected $x_{i,t}$), exploration is temporarily increased. This allows the agent to re-evaluate allocations under new conditions.
\end{enumerate}

Non-stationarity and noisy rewards can destabilize RL training. To improve stability, we incorporate the following techniques:
\begin{itemize}
    \item \textbf{Adaptive learning rates.} Gradually decaying $\alpha$ ensures convergence once probabilities stabilize, reducing oscillations.
    \item \textbf{Regularization.} Penalties on large Q-value updates prevent instability under abrupt shifts \cite{farahmand2011}.
    \item \textbf{Reward smoothing.} Using rolling averages of rewards reduces variance caused by transient fluctuations.
\end{itemize}

\section{Simulation Study}
\label{sec:simulation}

To evaluate the effectiveness of the proposed reinforcement learning framework, we conduct a simulation study designed to emulate heterogeneous pre-production testing environments with dynamic and non-stationary properties. Our objective is to compare the RL-based allocation strategy with both static and optimization-based baselines, under realistic conditions where configuration failure probabilities evolve over time. Simulation provides a controlled setting where ground-truth probabilities are known, allowing us to benchmark performance against an oracle that always allocates optimally with respect to true probabilities.

\subsection{Simulation Setup}

We consider $N=300$ testing units that must be allocated across $C=10$ configuration types over a horizon of $T=100$ discrete time steps. At each time step $t$, the agent selects an allocation $\{n_1(t), \ldots, n_C(t)\}$ subject to the budget constraint $\sum_{i=1}^C n_i(t)=N$. Each configuration type $c_i$ has a time-varying probability $p_i(t)$ of producing a detectable failure signal. Given allocation $n_i(t)$, the number of observed signals is
\[
X_i(t) \sim \text{Binomial}\!\big(n_i(t), p_i(t)\big).
\]
The key performance metric is the coverage measure
\[
D_t = \sum_{i=1}^C \mathbf{1}\{X_i(t)\geq 1\},
\]
representing the number of configuration types that produce at least one failure signal at time $t$. Maximizing $D_t$ reflects the goal of broad coverage across configurations.

To mimic realistic environments, we allow detection probabilities $\{p_i(t)\}$ to vary over time. Rather than fixed values, each $p_i(t)$ follows a stochastic process that includes both gradual fluctuations and abrupt shifts. Specifically, we simulate the complements $q_i(t)=1-p_i(t)$ as follows:

\begin{itemize}
    \item \textbf{Initialization:} For each configuration type $c_i$, set $q_i(0)\sim \text{Beta}(6,1)$ to initialize probabilities near zero (most systems are stable under nominal conditions).
    \item \textbf{Stochastic drift:} For each $t>0$, update
    \[
    q_i(t) = \text{clip}\!\big(q_i(t-1) + \varepsilon_{i,t}, 0, 1\big), \quad \varepsilon_{i,t}\sim \mathcal{N}(0,\sigma^2),
    \]
    with $\sigma^2$ tuned to produce modest temporal variability.
    \item \textbf{Abrupt shifts:} At pre-specified time points, selected types experience regime changes (e.g., sudden increase in failure probability due to latent bugs). For example:
    \begin{itemize}
        \item $c_1$: $q_1(t)$ decreases from $\approx 0.9$ to $\approx 0.7$ at $t=30$.
        \item $c_2$: $q_2(t)$ increases from $\approx 0.7$ to $\approx 0.95$ at $t=40$.
        \item $c_3$: $q_3(t)$ increases from $\approx 0.8$ to $\approx 0.95$ at $t=50$.
    \end{itemize}
    Other types retain their initial levels with only minor fluctuations.
\end{itemize}

\begin{figure}[h]
    \centering
    \includegraphics[width=\linewidth]{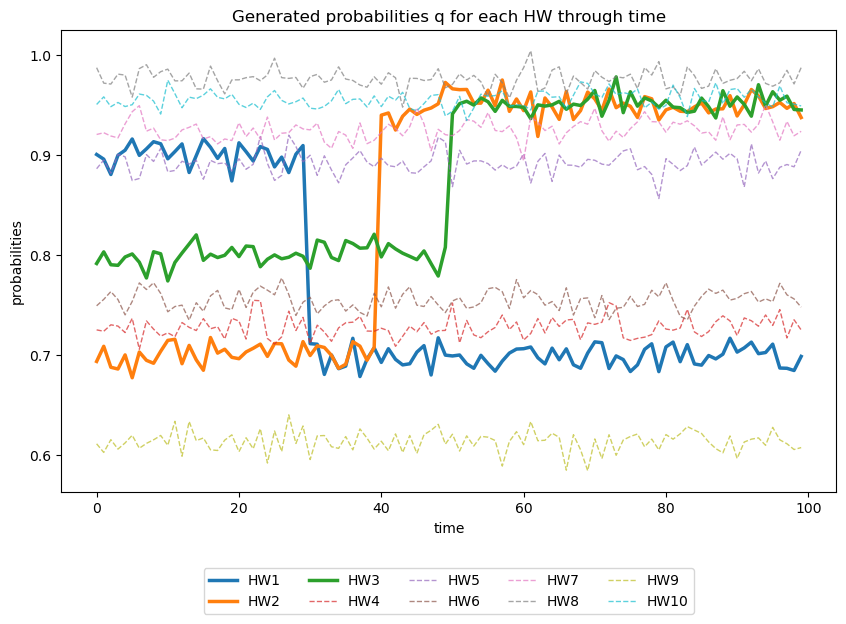}
    \caption{Example realization of configuration failure probabilities $p_i(t)$ over time. Several types undergo abrupt regime changes, creating non-stationary dynamics that challenge static allocation strategies.}
    \label{fig:prob_over_time}
\end{figure}

This process generates realistic non-stationary dynamics, similar to those observed in adaptive experiment design, bandit problems with drifting rewards, and non-stationary RL benchmarks \cite{garivier2011, besbes2014}. Figure~\ref{fig:prob_over_time} shows one realization of the simulated probabilities.

In real-world testing, true probabilities $p_i(t)$ are unknown and must be estimated from observed signals. We therefore design the baselines and RL agent to rely only on empirical estimates. After an initial burn-in period of $L=10$ time steps, each $\hat{p}_i(t)$ is estimated using a rolling weighted average of observed failure rates:
\[
\hat{p}_i(t) = \sum_{k=1}^{L} \omega_k \frac{X_i(t-k)}{n_i(t-k)}, \quad \omega_k \propto \frac{1}{k}.
\]
To prevent degeneracy, estimates are clipped to lie in $[\epsilon, 1-\epsilon]$ with $\epsilon=10^{-6}$. This estimator favors recent information, allowing the system to track gradual drift.

We mainly compare between four allocation strategies:

\begin{enumerate}
    \item \textbf{Static Baseline.} After burn-in, an allocation is optimized once using the estimated probabilities $\hat{p}_i$ from the initial window and then held fixed for all subsequent time steps. This reflects common industrial practice where test allocations are optimized periodically but not updated dynamically.

    \item \textbf{Rolling Lagrangian Method.} At each $t>L$, a new allocation is computed by solving a Lagrangian-threshold optimization using the latest $\hat{p}_i(t)$. This method adapts gradually as estimates evolve, but does not explicitly model sequential decision-making.

    \item \textbf{RL with Q-learning.} The proposed RL agent allocates adaptively using the Q-learning framework described in Section~\ref{sec:methodology}. Rewards are shaped by combining simulated outcomes (using $\hat{p}_i(t)$) with observed outcomes. Offline simulations are used to augment learning, analogous to experience replay \cite{mnih2015}.

    \item \textbf{Oracle.} For benchmarking, we compute the optimal allocation at each $t$ using the true probabilities $p_i(t)$. This strategy is unattainable in practice but provides an upper bound on achievable performance.
\end{enumerate}

\subsection{Evaluation}

We report results using two complementary metrics:

\begin{itemize}
    \item \textbf{Coverage ($D_t$).} The number of configuration types with at least one detected signal. Higher values indicate broader representation across configurations.
    \item \textbf{Mean Squared Error (MSE).} The squared deviation between estimated $\hat{p}_i(t)$ and true $p_i(t)$, averaged over types. Lower values indicate more accurate estimation and more stable allocation.
\end{itemize}

These metrics jointly capture both allocation quality and the agent’s ability to track probability dynamics.

\begin{figure*}[!h]
    \centering
    \includegraphics[width=0.8\textwidth]{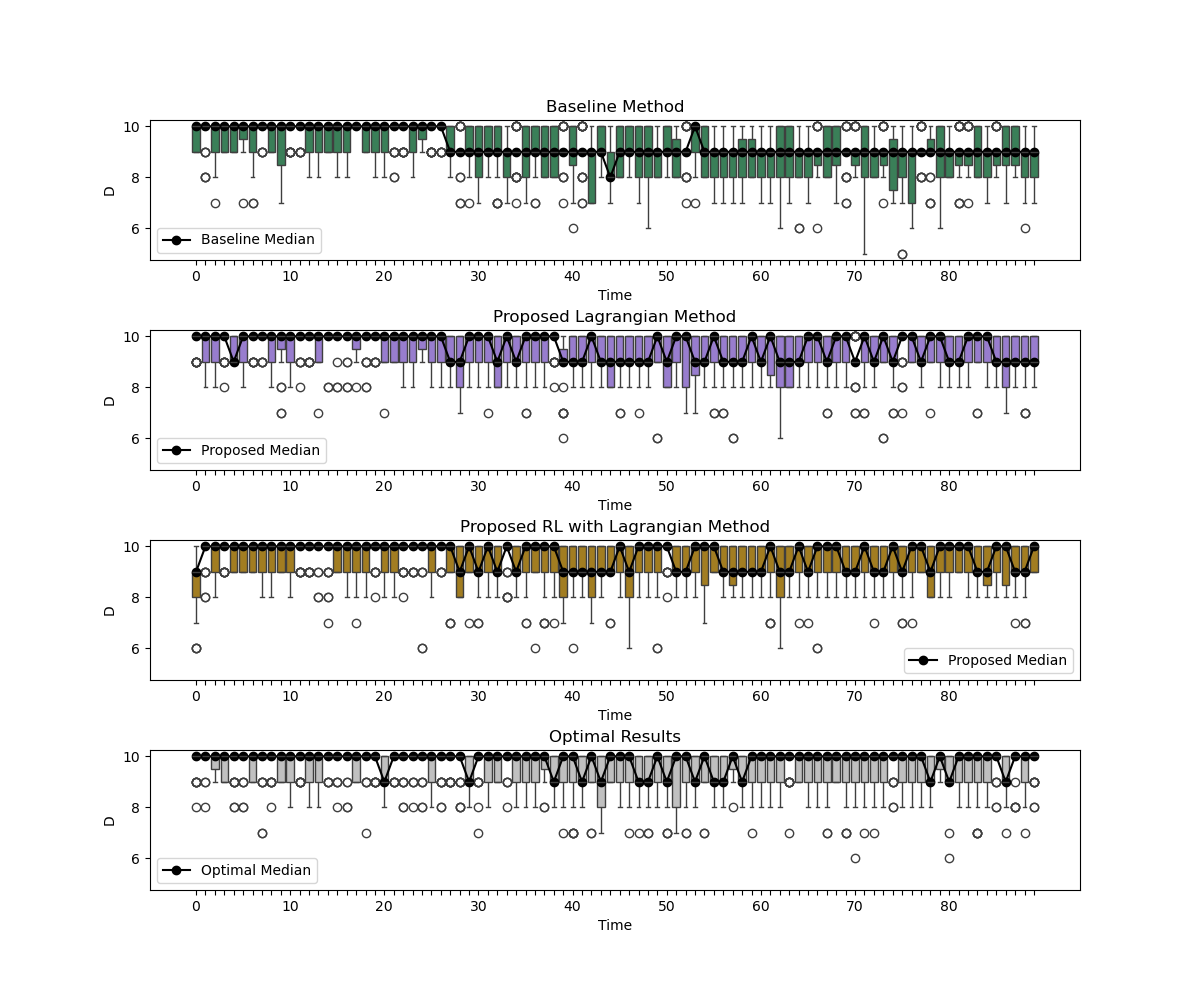}
    \caption{Coverage metric $D_t$ across $T=100$ steps for baseline, rolling Lagrangian, RL, and oracle strategies. RL adapts effectively to abrupt shifts, maintaining near-oracle coverage.}
    \label{fig:Dt_RL_over_time}
\end{figure*}

\begin{figure}[!h]
    \centering
    \includegraphics[width=\linewidth]{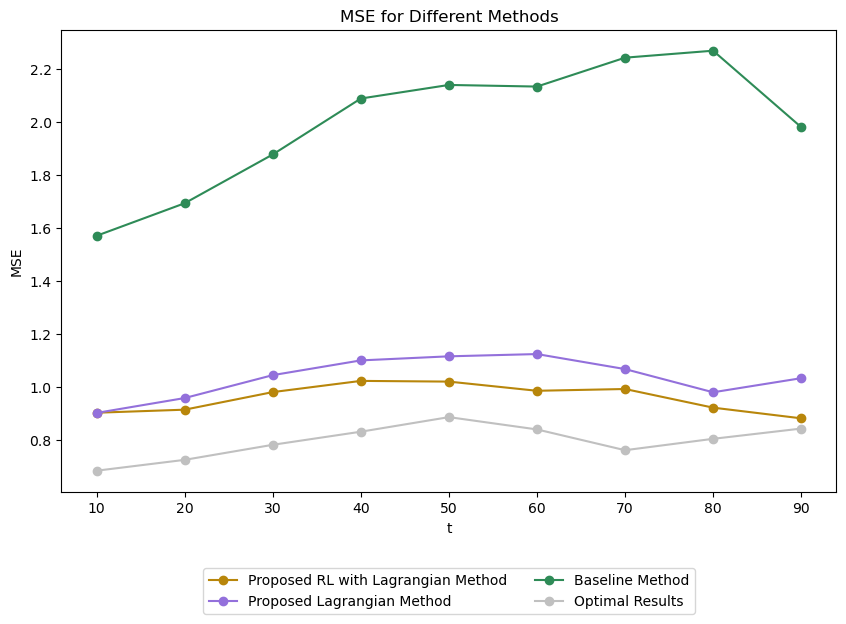}
    \caption{Mean squared error (MSE) of estimated probabilities across strategies. The RL agent achieves lower long-run error than the static baseline and performs comparably to the rolling Lagrangian method.}
    \label{fig:MSE_over_time}
\end{figure}

We repeat the simulation for $n_{\text{sims}}=50$ runs to assess robustness. Figure~\ref{fig:Dt_RL_over_time} plots the coverage metric $D_t$ across methods, while Figure~\ref{fig:MSE_over_time} compares MSE trajectories.

The static baseline performs poorly after regime shifts, as its fixed allocations become mismatched to evolving probabilities. The rolling Lagrangian method performs better by updating estimates but remains limited by lag in adaptation. The RL approach exhibits superior adaptability, recovering quickly after abrupt shifts and sustaining higher coverage, with performance close to the oracle benchmark. These results demonstrate the advantage of sequential decision-making in dynamic environments.

To quantify differences between adaptive methods, we apply the Wilcoxon signed-rank test \cite{wilcoxon1945, pratt1959, conover1999} to paired results from the rolling Lagrangian and RL strategies across all time steps and replications. Let $X_{t,i}$ and $Y_{t,i}$ denote coverage values under Lagrangian and RL, respectively. The null hypothesis is that the median difference is zero:
\[
H_0: \text{median}(X_{t,i}-Y_{t,i})=0.
\]
The test statistic is
\[
W = \sum_{t,i} R\!\big(|D_{t,i}|\big) \cdot \text{sign}(D_{t,i}),
\quad D_{t,i} = X_{t,i}-Y_{t,i}.
\]
Results indicate a statistically significant improvement of the RL method over the rolling Lagrangian approach (p-value $<0.05$), confirming that the performance gains are robust across simulations.

\section{Conclusion}
\label{sec:conclusion}

In this paper, we addressed the problem of adaptive configuration allocation in pre-production testing environments, where limited testing resources must be distributed across heterogeneous and dynamic system configurations. We highlighted the limitations of traditional combinatorial optimization approaches, which rely on static assumptions, ad hoc parameterization, and limited integration of real-time feedback. To overcome these challenges, we proposed a reinforcement learning framework based on Q-learning, which treats allocation as a sequential decision-making process.  

Our methodology introduced several key innovations: (i) formalizing the allocation task as an RL problem with explicit state, action, and reward definitions; (ii) designing a hybrid reward-shaping mechanism that integrates simulated and observed outcomes to reduce sample complexity; and (iii) incorporating online and offline updates to adapt effectively to non-stationary environments. These features collectively allow the agent to balance exploration and exploitation, maintain stability under regime shifts, and approach near-optimal allocation performance.  

Through a series of controlled simulation experiments, we compared the RL-based approach with static and optimization-based baselines. Results showed that the static baseline struggled under dynamic conditions, while the rolling Lagrangian method improved adaptation but lagged behind during abrupt changes. The proposed RL agent consistently achieved higher coverage and lower estimation error, closely approximating the oracle strategy that has access to true probabilities. Statistical analysis using the Wilcoxon signed-rank test confirmed the robustness of these improvements across replications.  

Our work contributes to the broader literature on adaptive testing, non-stationary reinforcement learning, and combinatorial optimization. It demonstrates that RL can provide a principled and effective solution for real-world evaluation pipelines where heterogeneity and non-stationarity are intrinsic. Beyond pre-production testing, the framework is relevant to applications in adaptive experiment design, dynamic resource allocation \cite{mao2016}, and non-stationary multi-armed bandits \cite{besbes2014, yu2020}.  

Several limitations suggest directions for future work. First, our study focused on discrete Q-learning with modest configuration spaces; scaling to larger and higher-dimensional environments may require deep reinforcement learning \cite{mnih2015} or policy-gradient methods. Second, our simulations assumed structured but synthetic probability dynamics; validating the approach on real-world testing pipelines would provide stronger evidence of practical effectiveness. Third, the current formulation optimizes coverage as the primary metric, but in practice, multi-objective trade-offs (e.g., failure severity, testing cost, latency) must also be considered. Extending the framework to multi-objective or constrained RL settings \cite{altman1999, achiam2017} is a promising direction.

In summary, reinforcement learning provides a powerful paradigm for adaptive configuration allocation in heterogeneous testing environments. By systematically integrating feedback, handling non-stationarity, and balancing exploration with exploitation, RL-based approaches can enhance the robustness and statistical power of pre-production testing, ultimately improving the reliability and stability of deployed systems.

\newpage
\section*{Appendices}

\subsection*{(a). Lagrangian Method}

In this section, we describe the problem settings and the approach used to solve the allocation problem using Lagrangian optimization. Let \( q_i \) denote the probabilities of an event not being raised (i.e., the probability of not catching the signal), \( n_i \) represent the allocations for each probability \( q_i \), and \( N \) be the total number of allocations.

The objective is to allocate \( N \) across different \( n_i \) such that the sum of \( n_i \) equals \( N \). This can be formulated as an optimization problem where the objective function is minimized subject to the constraint:

\begin{equation}
\sum_{i} n_i = N.
\end{equation}

\subsection*{(b). Derivation of the Function \( f \)}

The Lagrangian function for this problem is given by:

\begin{equation}
\mathcal{L}(n_i, \lambda) = \sum_{i} g(n_i, q_i) - \lambda \left( \sum_{i} n_i - N \right),
\end{equation}

where \( g(n_i, q_i) \) is the probability \( P(J_i < \tau) \) where \( J_i \sim \text{Bin}(n_i, q_i) \), and \(\lambda\) is the Lagrange multiplier.

The probability \( P(J_i < \tau) \) is defined as:

\begin{equation}
P(J_i < \tau) = \sum_{j=0}^{\tau-1} \binom{n_i}{j} (1 - q_i)^j q_i^{n_i - j} ,
\end{equation}

where \( \binom{n_i}{j} \) is the binomial coefficient. For different thresholds \(\tau\), this function varies as follows:

\begin{itemize}
    \item \(\tau = 1\):
    \begin{equation}
    g(n_i, q_i) = P(J_i < 1) = q_i^{n_i}
    \end{equation}
    \item \(\tau = 2\):
    \begin{equation}
    g(n_i, q_i) = P(J_i < 2) = q_i^{n_i} + n_i q_i^{n_i - 1} (1 - q_i)
    \end{equation}
    \item \(\tau = 3\):
    \begin{equation}
    g(n_i, q_i) = P(J_i < 3) = q_i^{n_i} + n_i q_i^{n_i - 1} (1 - q_i) + \frac{n_i (n_i - 1)}{2} q_i^{n_i - 2} (1 - q_i)^2
    \end{equation}
\end{itemize}

To find the optimal \( n_i \), we derive the partial derivative of the Lagrangian with respect to \( n_i \) and set it to zero. 
For example, for \(\tau = 1\), the closed form of \( n_i \) can be obtained directly from:
\begin{equation}
\frac{\partial }{\partial n_i} (q_i^{n_i} - \lambda n_i) = 0
\end{equation}

which gives:

\begin{equation}
n_i = \frac{\log\left(\frac{\lambda}{\log(q_i)}\right)}{\log(q_i)}.
\end{equation}

When  \(\tau = 3\), the derivative function \( f \) is derived as follows:

\begin{equation}
f(n_i, q_i, \lambda) = q_i^{n_i} \left[ \log(q_i) \left( 1 + n_i \frac{1 - q_i}{q_i} + \frac{n_i (n_i - 1)}{2} \left( \frac{1 - q_i}{q_i} \right)^2 \right) + \left( \frac{1 - q_i}{q_i} + \frac{(1 - q_i)^2}{2 q_i^2} (2n_i - 1) \right) \right] - \lambda
\end{equation}

In this case, we would need to apply numerical analysis to solve $f(n_i, q_i, \lambda) = 0$.

\subsection*{(c). Algorithm to Find Optimal \(\lambda\) and \( n_i \)}

The algorithm to find the optimal \(\lambda\) and corresponding \( n_i \) involves the following steps \ref{alg:optimal_lambda}:

\begin{algorithm}[!h]
\caption{Lagrangian algorithm to find optimal \(\lambda\) and \( n_i \)}
\label{alg:optimal_lambda}
\begin{algorithmic}[1]
\STATE Initialize the probability array \( q_i \) and total number of allocations \( N \).
\STATE Define the range for \(\lambda\), \(\lambda_{\text{min}}\) and \(\lambda_{\text{max}}\), and set the number of points for the grid search.
\STATE Define the tolerance \( \epsilon \) and maximum iterations for the bisection method.
\FOR{each \(\lambda\) in the grid search range}
    \FOR{each \( q_i \)}
        \IF{\(\tau = 1\)}
            \STATE Obtain \(n_i\) with closed form solution.
        \ELSE
            \STATE Obtain \(n_i\) using the bisection method to solve \( f(n_i, q_i, \lambda) = 0 \).
        \ENDIF
    \ENDFOR
    \STATE Calculate the sum of \( n_i \) values.
    \IF{the constraint \( \left| \sum_i n_i - N \right| < \epsilon \)}
        \STATE Compute the objective function.
        \STATE Update the optimal \(\lambda\) if the current objective function value is lower.
    \ENDIF
\ENDFOR
\STATE \textbf{return} the optimal \(\lambda\) and corresponding \( n_i \).
\end{algorithmic}
\end{algorithm}

\subsection*{(d). Bisection Method}

The bisection method used in the algorithm is as follows \ref{alg:bisection_method}:

\begin{algorithm}[!h]
\caption{Bisection Method}
\label{alg:bisection_method}
\begin{algorithmic}[1]
\STATE Initialize \( n_{i1} \) and \( n_{i2} \) such that \( f(n_{i1}) \) and \( f(n_{i2}) \) have opposite signs.
\FOR{each iteration until convergence or maximum iterations}
    \STATE Calculate \( n_i = \frac{n_{i1} + n_{i2}}{2} \).
    \STATE Evaluate \( f(n_i, q_i, \lambda) \).
    \IF{\( |f(n_i, q_i, \lambda)| < \epsilon \)}
        \STATE Return \( n_i \).
    \ELSIF{\( f(n_{i1}, q_i, \lambda) \cdot f(n_i, q_i, \lambda) < 0 \)}
        \STATE Set \( n_{i2} = n_i \).
    \ELSE
        \STATE Set \( n_{i1} = n_i \).
    \ENDIF
\ENDFOR
\STATE Raise an error if the solution does not converge.
\end{algorithmic}
\end{algorithm}

\bibliographystyle{ACM-Reference-Format}
\bibliography{sample-base}

%%%%%%%%%%%%%%%%%% CASE STUDY %%%%%%%%%

\end{document}